\documentstyle[12pt,aasms4]{article}

\def\ltsima{$\;\buildrel < \over \sim \;$}
\def\simlt{\lower.5ex \hbox{\ltsima}}
\def\gtsima{$\;\buildrel > \over \sim \;$}
\def\simgt{\lower.5ex \hbox{\gtsima}}
\begin{document}
\title{The Cosmic Infrared Background Radiation, Star Formation Rate,  and Metallicity}

\author{Martin Harwit}

\vskip 0.5 true in \parskip 6pt

\noindent{511 H Street S.W., Washington, DC 20024--2725; also Cornell University}

\vskip 0.2 true in


\begin{abstract}
The density of the far infrared / submillimeter (FIR/SMM) diffuse extragalactic radiation field has recently been
determined from COBE data. Nearly simultaneously, deep FIR/SMM  surveys
have detected substantial numbers of optically unidentified sources, which have led to the
proposal that galaxies and protogalaxies at red shifts $ z = 2$ to 4  may account for an
appreciable fraction of the background.  Here I show that, if the reported radiation levels are
generated through nucleosynthesis,  most of this energy must have been produced at epochs $z
\lesssim 2$. Hubble Deep Field data cited by Madau et al.\ (1998) indicate that the bulk of the
integrated extragalactic background must have been generated even more recently,  at $z < 1$.
\end{abstract}

\section{Introduction}

Two critical observational advances have taken place in the past two years.  The first is the
determination of a diffuse extragalactic FIR/SMM background (Puget et al.\ 1996; Hauser et al.\ 1998).  The second is the detection of faint distant galaxies at FIR/SMM wavelengths (Kawara et al.\
1998; Hughes, et al.\ 1998; Barger et al.\ 1998).   Hughes et al., who found no visible or infrared galaxies at the
positions of most of the sources they reported in the Hubble Deep Field, suggest that these and much of the SMM background originate in distant, dust-shrouded galaxies at red shifts $z = 2$ to 4, undetectable at short
wavelengths. 

A number of observations argue against such an early origin for most of the diffuse background.  Lanzetta et al.\ (1995) have
suggested that most Ly-$\alpha$ absorption systems are progenitors of galaxies, hydrogen rich at high red shifts and rich in stars at low red shifts. This is supported by Rao, Turnshek, and Briggs (1995), who found a drop in gas content by an order of magnitude between $z\sim 3.5\ {\rm and}\ 0$, but roughly comparable density parameters, for gas at high red shifts, $\Omega_g (z = 3.5)$, and luminous matter (stars) seen today, $\Omega_s(z = 0)$. 

Pettini et al.\ (1998) have traced the evolving abundance of zinc in damped Ly$\alpha$ systems from red shifts $z = 3.4\ {\rm to}\ 0.4$.  Zinc is a
useful tracer of metallicity, since it is not readily deposited from the gaseous phase onto interstellar grains.  At red shifts beyond $z \sim 2.8$ only upper limits  could be established, indicating a zinc deficiency $>25$. But even at red shifts $z = 1$ the zinc abundance remains a factor of $\sim 3\ {\rm to}\ 10$ below the solar abundance.  At red shifts $z\gtrsim 3$, the abundance of carbon in these systems is roughly a factor of 100 lower than solar, and independent of the optical depth (Songaila \& Cowie\ 1996; Cowie \& Songaila\ 1998).  These results do not appear to be seriously affected by dust absorption, since the extinction even for blue light reaches unit optical depth only rarely, at the highest column densities, $\sim 2\ {\rm to}\ 3\times 10^{21}\,{\rm cm}^{-2}$ (Zuo et al. 1997, Songaila et al. 1995).  The cited abundances of heavy elements thus are likely to be reasonable tracers of energy generation in massive and supermassive stars at early epochs.  These metallicities also agree with data for the oldest Galactic stars formed at about the same epoch (Timmes et al.\ 1995; Cowan et al.\ 1995).

The general picture that emerges is that the earliest production of heavy elements occurred in Population III stars whose masses were sufficiently high to produce and explosively eject r-process elements.  The metallicity at these  epochs was 2 to 3 orders of magnitude lower than today.  Substantial accumulations of heavy elements did not appear until $z \sim 2$.   
 
Sections 2 and 3, below, respectively summarize the background observations and FIR/SMM
surveys.  Section 4 discusses energy generation rates, while section 5 shows why the primary epoch of energy generation must lie in the relatively recent past, at red shifts $z\lesssim 2$.

\section{Background Observations}

Using the Far-Infrared Absolute
Spectrometer (FIRAS) onboard the Cosmic Microwave Explorer satellite (COBE), Guideroni, et al.\ (1997) estimated a flux at wavelengths between 200 and 400\,$\mu$m, of $\lambda I_{\lambda}  \sim 7$\,nW\,m$^{-2}\,$sr$^{-1}$ at $\sim 300\mu$m, corresponding to a radiation density $\rho_{300\mu{\rm m}}\sim 3\times 10^{-15}$\,erg cm$^{-3}$.  Measurements from the Diffuse Infrared Background
Experiment (DIRBE) aboard the same satellite led Hauser et al.\ (1998) to a somewhat higher integrated energy from 140 to 240 $\mu$m, of $\lambda I_{\lambda}\sim 10\,$nW\,m$^{-2}\,$sr$^{-1}$, or $\rho_{200\mu{\rm m}}\sim 4\pi \lambda I_{\lambda}/c \sim 4\times 10^{-15}$ erg cm$^{-3}$.  Fixsen et al.\ (1998) used both DIRBE and FIRAS data and found the total FIR/SMM background from 125$\mu$m to 2\,mm to be slightly higher, $\rho_{FIR/SMM}\sim 6\times 10^{-15}$ erg cm$^{-3}$.  

These data represent only a fraction
of the total diffuse extragalactic energy density.  In the mid-infrared, between $\sim 5$ and 100$\,\mu$m, only coarse upper limits are available, all of which are much higher than the FIR/SMM values.  At 3.5$\,\mu$m, Dwek \& Arendt (1998) obtain a radiation density of $3\times 10^{-15}$ erg\,cm$^{-3}$ from DIRBE observations.  At 3600 to 22,000\,\AA, Pozzetti et al\ (1998) have estimated a background flux of $\sim 4\times 10^{-15}$ erg cm$^{-3}$, from discernable sources in the Hubble Deep Field.  We can write the total extragalactic radiation density,
exclusive of the 2.73\,K microwave background radiation, as  
\begin{equation}
\rho_{\nu}\sim \frac{6\times 10^{-15}}{f_t}\ {\rm erg\ cm}^{-3}
\end{equation}
where $0.05\leq f_t\leq 0.5$ is the fraction that lies in the FIR/SMM range. The upper limit to $f_t$ comes from the UV/optical/near-infrared contribution; the lower limit comes from  uncertainties, comprehensively reviewed by Dwek et al. (1998), and from the implications of the observed TeV gamma-ray flux from Mrk 501 (Coppi \& Aharonian 1997; Stanev \& Franceschini 1998).

\section{Contributions from Discrete Sources}

Using the Infrared Space Observatory, ISO, Taniguchi et al.\ (1997) found 15 sources at 5 to 8.5$\,\mu$m, with a total flux of 1\,mJy, in a $3'\times 3'$ field, corresponding to a radiation density of $\rho_{7\mu{\rm m}}\sim 1.5\times
10^{-16}$\,erg\,cm$^{-3}$.  At 12 to 18$\mu$m, Altieri et al.\ (1998) derived a preliminary density of sources at fluxes above 50$\,\mu$Jy at $2.5\times 10^{-4}$ per square degree, yielding $\rho_{15\mu{\rm m}}\gtrsim 1.5\times 10^{-16}$\,erg\,cm$^{-3}$ as well. Kawara et al.\ (1998) conducted a deep survey over a 1.1 square degree field at 95 and 175$\,\mu$m, respectively, and found 36 and 45 sources,  all brighter than 150\,mJy.  The corresponding radiation density in this $\sim 140\,\mu$m-wide band at 120$\,\mu$m is 
$\rho_{120\mu{\rm m}}\gtrsim 1.5\times 10^{-16}\,$erg cm$^{-3}$.

At 850$\,\mu$m two deep surveys have been undertaken, both using the SCUBA array  on the James Clerk Maxwell telescope in Hawaii.  Hughes
et
al.\ (1998) reported five sources in a 5.6 square arc minute field for a total flux of 20\,mJy, or $\lambda I_{\lambda} = 1.5 \times10^{-10}\,$W\,m $^{-2}$\,sr$^{-1}$, while Barger et al.\ (1998) have found two sources with a combined flux of 8 mJy in two such fields of view.  The
respective energy densities at 850\,$\mu$m are $\sim 6\times
10^{-17}$ and $\sim 1.3\times 10^{-17}$\,erg\,cm$^{-3}$.  

\section{Star Formation Rates, Metallicity, and Energy Production}

Madau et al.\ (1998) have made a persuasive case for estimating early star formation and energy generation rates, on the basis of luminosity of distant sources in the 0.15 to 2.2 $\mu$m wavelength band.  The top curve in Figure 1  shows their derived energy generation rates as a function of red shift.  This approach is complementary to estimates of energy generation rates based
on observed metallicities (Fall et al.\ 1996). 

Directly detected metallicity can underestimate the true heavy element and energy production.  The fraction of processed material ejected from stars, $f_{ej}$, and the energy per unit mass, $\varepsilon_{ej}$, generated in the process are only part of the story.  We need to also consider the fraction of all baryonic mass, $f(z)$, which has radiated away energy in producing heavy elements retained in the interior of stars. 

     a.   White dwarfs constitute a fraction $f_{wd}\sim 0.1$ of the mass of all stars.  Vassiliadis and Wood\ (1993,\ 1994) have traced the evolution of stars with initial masses ranging from 0.89 to 5 M$_{\odot}$, and find that the ejected  mass contains hardly any elements heavier than helium, that the helium content of the ejected gas is only a few percent higher than the initial fraction, but
that the final core masses of processed material range from $\sim 0.55\ {\rm for \ the \ lower\ mass\ stars\ to}\, \sim 0.9\,$M$_{\odot}$ for stars at the high end of the range.  We may, therefore, approximate the energy that was generated per unit white dwarf mass as a fraction $\varepsilon_{wd}\sim 0.008$ of the star's current mass.  

b.  Neutron stars constitute a fraction $f_{ns}\sim 0.01$ of all stellar mass, and have also
converted a fraction $\varepsilon\sim 0.008$ of this mass into energy. The neutron star's rotational energy immediately after collapse into a body spinning with a millisecond period is also of order $0.2\%$ of the mass-energy. This is eventually dissipated and radiated away.  A typical neutron star with mass $M\sim  1.4\,$M$_{\odot}$, thus will radiate
away
a fractional amount of energy per unit mass $\varepsilon_{ns}\sim 0.01$.  

c.   The supernova explosion accompanying the formation of a neutron star  
provides the heavy elements and much of the non-primordial helium observed in
stellar atmospheres and the interstellar medium.  Chieffi et al.\ (1998) have recently compared estimates by various authors, of the total mass of heavy-element ejecta from supernovae, as contrasted to mass retained in the core.  For a star whose initial mass was 25M$_{\odot}$ the current consensus appears to favor ejection of a mass of approximately 4M$_{\odot}$ in carbon,
oxygen, neon, magnesium and silicon.  In addition, the star also yields 9\,M$_{\odot}$ of $^4$He, and a core mass of iron M$_{Fe}\sim 1.5\,$M$_{\odot}$.   Consistent with the ratio of heavier elements to helium in supernova ejecta and stellar winds, we find that heavy elements constitute about 2\% of the sun's mass, and that the sun's helium content is roughly 30\% by mass, about 6\% higher than the primordial helium abundance.  The ratio of heavy elements ejected to those retained in neutron stars is also consistent with an overall neutron star complement amounting to $\sim$\,1\% of all stellar mass. 
We write $f_{ej}\sim 0.02,\ \varepsilon_{ej} \sim 0.01$, respectively determined by solar heavy-element abundance and nuclear plus kinetic energies released in supernova explosions.

d.  Stellar mass black holes may radiate away an amount of energy comparable to neutron stars, but they appear to be rare, and probably do not contribute greatly to the overall electromagnetic energy constituting the extragalactic background.  Neither the fractional mass in black holes at the centers of galaxies, nor the efficiency with which they have radiated away electromagnetic energy are currently well established, but their contribution to the metallicity and electromagnetic background appears to be modest.

e. Finally, we need to also consider the class of low-mass stars that have never reached the white
dwarf stage, but have nevertheless contributed to the radiation density, though not to the
observed metallicity. Their fraction by mass, $f_{\ell m}$, depends on the stellar birth rate
function.

These different sources of energy suggest three separate estimates of contributions to the
extragalactic background.  The first is for a single burst, at red shift $z$, of rapidly evolving,
massive, primordial, Population III objects with a fractional mass   $f_{ej}(z) +
f_{ns}(z)$.   The
second is for a continuous generation of massive stars that evolve primarily into neutron
stars and white dwarfs. Here, the rate of formation at epoch $z$ is $\dot f(z) = \dot
f_{ej}(z)
+ \dot f_{ns}(z) + \dot f_{wd}(z)$, which needs to be integrated over the appropriate red-shift
interval $\Delta z$.
The last is for low-mass stars that may have formed at early times and have continued to shine at
roughly constant luminosity for most of the history of the Universe.  These cases are then confronted with current data.

\section{Must Most of the Energy Production have Occurred at Low Red Shifts $z$?}

The contribution to the radiation field by massive stars can be estimated from the dependence of the observed metallicity $Z$ on red shift $z$.  To lowest order, the data cited in Section 1 are consistent with a systematic metallicity decline by a factor of 10 for increasing red-shift intervals $\Delta z = 2$ 
\begin{equation}
 Z/Z_{\odot} \sim 10^{-z/2} \sim e^{-1.15z}
\end{equation}

\subsection{A Single Star Burst at Red Shift $z$}

 At epoch $z$, the baryonic mass density $\rho_B$ in a flat universe with
deceleration parameter $q_o = 0.5$ is 
\begin{equation}
\rho_B \sim \frac{3H_0^2 (1 + z)^3}{8\pi G}\Omega_B
\end{equation}
Here, $\Omega_B$ is the baryon density parameter, $H_o$ is the Hubble constant today, and
$G$ is the gravitational constant.  The fraction of mass that has been converted into heavy
elements by epoch $z$, and the energy generated in this process is given by the stars that
complete their evolution on time scales short compared to the age of the Universe and end up as
supernova ejecta, neutron stars or black holes. We may approximate this by $f(z)= \alpha
10^{-z/2}$, where the value $\alpha \sim 0.1$ is based on a solar abundance of
heavy elements of $\sim 2\%$, non-primordial helium of $\sim 6\%$, and the complement of
neutron stars that currently constitute roughly 1\% of  baryonic  mass. Combining the nuclear and
kinetic energies cited in section 4, we may take $\varepsilon \sim 0.01$.

If all Population III objects were formed in a single burst at epoch $z$, the metallicity of equation (2) and the corresponding radiation density
$\rho_{\nu}$ observed today would be related by
\begin{equation}
\rho_{\nu}(z) = \biggl (\frac{3H_0^2(1+z)^3\Omega_B}{8\pi G}\biggr )\biggl ( \alpha 10^{-z/2}
\biggr )\frac{c^2\varepsilon}{(1+z)^4}
\end{equation}
or
\begin{equation}
\rho_{\nu}(z) = 1.5\times 10^{-15}\biggl (\frac{H_0}{50\,{\rm km\
s}^{-1}\,{\rm Mpc}^{-1}}
\biggr )^2\biggl
(\frac{\Omega_B}{0.01} \biggr )\biggl (\frac{\alpha}{0.1} \biggr )\biggl
(\frac{\varepsilon}{0.01}\biggr )\biggl (\frac{3}{1+z}\biggr )10^{(2-z)/2}\ {\rm erg\ cm}^{-3}
\end{equation}
Comparison to equation (1) shows that a single burst, even as late as $z = 2$, fails to provide the requisite energy density for a value of $\Omega_B$ as high as $\sim 0.08$.   

\subsection{Continuous Formation of Massive Stars}

For continuous formation of massive stars from the earliest epochs, we note that the energy
production rate is just the derivative of $f(z)$
\begin{equation}
-\frac{df(z)}{dz}\varepsilon \sim 1.15\varepsilon\alpha_c e^{-1.15z}\sim 1.15\varepsilon\alpha_c
10^{-z/2}
\end{equation}
Here $\alpha_c \sim 0.2$, takes into account that roughly 10\% of the baryonic mass is in white dwarfs, and assumes that the birth rate function $\psi$ does not greatly change over the \ae ons. 
Integrating from $z = 5$ to the present, we have
\begin{equation}
\rho_{\nu}\sim 6\times 10^{-14}\biggl (\frac{H_0}{50\,{\rm km\ s}^{-1}\,{\rm Mpc}^{-1}}
\biggr )^2\biggl (\frac{\Omega_B}{0.01} \biggr )\biggl (\frac{\alpha_c}{0.2} \biggr )\biggl
(\frac{\varepsilon}{0.01}\biggr )\ {\rm erg\ cm}^{-3}
\end{equation}
This value is consistent with the observed background radiation even for $\Omega_B$ as
low as $\sim 0.003$, provided the FIR/SMM background is a
fraction $f_t\gtrsim 0.3$ of the total.  Since the metallicity rises sharply, all but $\sim 4\%$ of the light contributed to the background is of relatively
recent origin, $z \lesssim 2$.
\subsection{Low-Mass Stars}

For low-mass stars, we simplify the calculation by assuming that they all have mass  $M <
M_{\odot}$, were formed at a single epoch, and have steadily radiated at constant luminosity,
ever since. Taking a birth rate function $\psi \propto M$ and a luminosity proportional to $M^3$,
we find the mean luminosity for stars of mass $0.1M_{\odot} < M < M_{\odot}$ to be $\langle
L/M\rangle \sim 0.05L_{\odot}/M_{\odot}\sim 0.1\,{\rm erg\ (g\ s)}^{-1}$.  If a fraction $f_{\ell
m}$ of all the baryons is in low mass stars, the generated radiation energy density increases at a
rate
$\dot\rho_{\nu}(z)$, scaled down by the red shift. 
\begin{equation}
\rho _{\nu 0} = \int\frac{\dot\rho_{\nu}(z)}{(1+z)^4}dt = \int\langle L/M
\rangle\biggl(\frac{3H_0^2\Omega_B}{8\pi G} \biggr )
f_{\ell m} (1 + z)^{-1} dt = -\frac{3}{2}t_0\int\frac{\rho_{0B} f_{\ell m}} {(1+z)^{7/2}}dz
\end{equation}
where $\rho_{0B}$ is today's baryon density and the age of the Universe at any given time is
$t\sim t_0 (1+z)^{-3/2}$.  Stars that have been steadily shining since $z_{max} >>1$ contribute
\begin{equation}
\rho_{\nu 0} \sim 10^{-15}\biggl (\frac{\langle L/M \rangle }{0.1}\biggr )\biggl
(\frac{H_0}{50\,{\rm km\ s}^{-1}\,{\rm Mpc}^{-1}}\biggr )\biggl (\frac{f_{\ell
m}}{0.5}\biggr )\biggl (\frac{\Omega_B}{0.01}\biggr )\  {\rm erg\ cm}^{-3}
\end{equation}
to today's radiation density.  This is significantly lower than the
observed background. 

\subsection{Directly Observed Star Formation}

Finally, consider the observational star formation and energy generation
rate
given by Madau et al.\ (1998).  The top curve of Figure 1 shows their data for the UV energy generation rate in a comoving volume of 1 Mpc$^3$ in units of erg Hz$^{-1}$ s$^{-1}$ on the left-hand scale.  For the flat cosmological model with $q_0 = 0.5$ that they assume, the figure also shows the corresponding energy generation rate per unit red-shift interval in units of erg Hz$^{-1}$
z$^{-1}$ shown on the right.  Because time intervals $\Delta t$ corresponding to unit red-shift intervals, $\Delta z = 1$, diminish in proportion to $(1+z)^{5/2}$, this curve rapidly drops with increasing $z$.  The red shift further lowers the contribution to the background by $(1+z)$, for a cumulative reduction by $(1+z)^{7/2}$.  The integrated contribution to the background is $\sim 40$\,nW\,m$^{-2}$\,sr$^{-1}$ for an assumed UV/optical bandwidth $\Delta \nu \sim 3\times 10^{15}$\,Hz at the source.  Recent suggestions that dust absorption in early galaxies would raise the data at high red shifts, to flatten the curve beyond the peak at $z\sim 1.5$ (Hughes et al. 1998; Pascarelle et al. 1998) hardly change estimates of the background radiation received today.  Whatever the energy production rate at early times may be, the epochs at high red shifts do not last long, lead to an energy drain on the photons, and contribute little to the total energy density.   However, for $\alpha_c \sim 0.2$, a high, massive-star-formation rate at early epochs would require an increase in $\Omega_B$ to keep the neucleosynthetic helium and heavier element abundances from exceeding their observed values.

\section{Discussion}

The FIR/SMM extragalactic radiation density is most readily understood in terms of
energy generated by massive stars radiating primarily at recent epochs $z \leq 2$.  Conversely, for currently estimated baryon densities, massive stars at high red shifts could not have generated the high levels of diffuse infrared background radiation observed with COBE.  Almost any star
formation rate consistent with metallicities observed at different red shifts $z$ leads to the same
conclusion.  Energy generation rates could have been high at early epochs $z\gtrsim 4$, but the
duration of such epochs is always brief except in inflationary Lema\^{\i}tre universes, not
considered here.  

I thank Drs. Eli Dwek and Johannes Schmid-Burgk for important discussions and suggestions.   An anonymous referee suggested a number of significant improvements.  Dr.  Kawara and his colleagues permitted me to see their
paper before publication.  With pleasure I also acknowledge support provided by NASA grant NAG5-3347.  During the final stages of preparation of this paper, I was a guest of the Alexander von Humboldt Foundation of Germany and the Max Planck Institute for Radioastronomy in Bonn.  I thank both institutions for their warm hospitality.  

\vfill\eject
\centerline{\bf References}
\vskip 0.1 true in 
{\hoffset 20pt
\parindent = -20pt

Altieri, B., Metcalfe, L., Kneib, J.-P., and McBreen, B. 1998 ``Ultra-Deep Mid-IR Survey of a
Lensing Cluster," NGST Conference, Li\`ege, June 1998. 

Barger, A. J., et al. 1998, Nature, 394, 248

Chieffi, A., Limongi, M., \& Straniero, O. 1998, ApJ, 502, 737

Coppi, P. S. \& Aharonian, F. A. 1997, ApJ, 487, L9

Cowan, J. J., et al. 1995, ApJ, 439, L51

Cowie, L. L., \& Songaila, A. 1998, Nature, 394, 44

Dwek, E., et al. 1998, astro-ph/9806129

Dwek, E. \& Arendt, R. G. 1998, astro-ph/9809239

Fall, S. M., Charlot, S., \& Pei, Y. C. 1996, ApJ, 464, L43

Fixsen, D. J., et al. 1998, astro-ph/9803021

Guideroni, B., et al. 1997, Nature, 390, 257

Hauser, M. G., et al. 1998, astro-ph/9806167

Hughes, D. H., et al. 1998, Nature, 394, 241

Kawara, K., et al. 1998, A\&A, submitted 

Lanzetta, K. M., Wolfe, A. M., \& Turnshek, D. A. 1995, ApJ, 440, 435
 
Madau, P., Pozzetti L., and Dickinson, M. 1998, ApJ, 498, 106

Pascarelle, S. M., Lanzetta, K. M., \& Fern\'andez-Soto, A. 1998, astro-ph/9810060

Pettini, M., Ellison, S. L., Steidel, C. C., \& Bowen, D. V. 1998, astro-ph 9808017

Pozzetti, L., et al. 1998, MNRAS, 298, 1133

Puget, J.-L., et al. 1996, A\&A, 308, L5

Rao, S. M., Turnshek, D. A. \& Briggs, F. H. 1995, ApJ, 449, 488

Songaila, A., Hu, E. M., \& Cowie, L. L. 1995, Nature, 375, 124

Songaila, A., \& Cowie, L. L. 1996, AJ, 112, 335
 
Stanev, T. \& Franceschini, A. 1998, ApJ, 494, L159

Taniguchi, Y., et al.\ 1997, A\&A, 328, L9

Timmes, F. X., Lauroesch, J. T., and Truran, J. W. 1995, ApJ, 451, 468

Vassiliadis, E. \& Wood, P. R. 1993, ApJ, 413, 641

Vassiliadis, E. \& Wood, P. R. 1994, ApJS, 92, 125 

Zuo, L. et al. 1997, ApJ, 477, 568
}

\vfill\eject
\centerline{\bf Figure Caption}
\vskip 0.1 true in 
{\hoffset -20pt
\parindent -20pt
Fig. 1 -- Energy generation rates at different epochs, and their contribution to the present energy density.  The curve with symbols is from Madau et al.\ (1998) and refers to the comoving UV energy generation rate, per second, for a flat universe with zero pressure and a deceleration parameter $q_0 = 0.5$; here, the left-hand scale applies.  The right-hand scale refers to the comoving energy generation rate per unit interval $\Delta z$ and to today's received radiation density. The energy generation and reception rates in both these sets of units remain invariant under red shifts, but the bandwidth diminishes as (1+z) leading to a reduced effective reception rate.  These curves dramatically illustrate the negligible contribution that epochs at high red shifts make to the observed radiation density.  To match the assumptions of Madau et al., all curves assume $H_0 = 50$\,km\,s$^{-1}$\,Mpc$^{-1}$. }

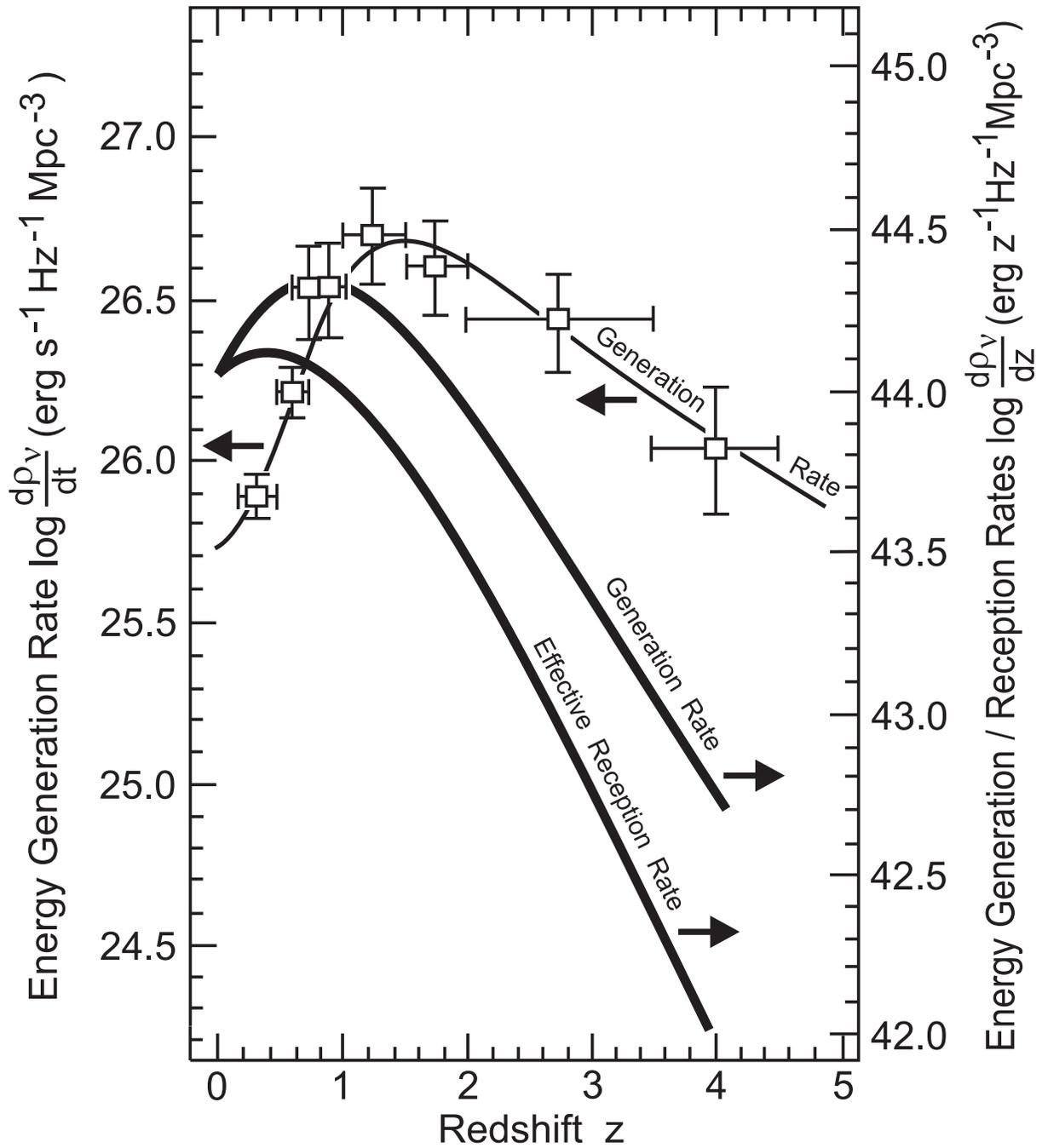
\end{document}